\documentclass{emulateapj}				

\slugcomment{Version of \today}                                               
\shortauthors{Harrington et al.}                                            
\shorttitle{Mueller Matrices Using Sky Polarization}

\begin{document}

\title{Deriving Telescope Mueller Matrices Using Daytime Sky Polarization Observations}

\author{David M. Harrington$^1$}

\author{J. R. Kuhn$^2$}

\author{Shannon Hall$^3$}

\affil{$^1$Institute for Astronomy, University of Hawaii, {2680 Woodlawn Drive}, Honolulu, HI, 96822}
\email{dmh@ifa.hawaii.edu}

\affil{$^2$Institute for Astronomy Maui, University of Hawaii, {34 Ohia Ku St.}, Pukalani, HI, 96768}
\email{kuhn@ifa.hawaii.edu}

\affil{$^3$Department of Astronomy, Whitman College, {345 Boyer Ave.}, Walla Walla, WA, 99362}
\email{hallsw@whitman.edu}

\begin{abstract}
Telescopes often modify the input polarization of a source so that the measured circular or linear output state of the optical signal can be signficantly different from the input. This mixing, or polarization ``cross-talk'', is defined by the optical system Mueller matrix. We describe here an efficient method for recovering the input polarization state of the light and the full $4\times 4$ Mueller matrix of the telescope with an accuracy of a few percent without external masks or telescope hardware modification. Observations of the bright, highly polarized daytime sky using the Haleakala 3.7m AEOS telescope and a coud\'{e} spectropolarimeter demonstrate the technique.

\end{abstract}

\section{Introduction}
	
	Spectropolarimetry is a powerful tool which may be enhanced by efficient techniques for telescope-detector polarization calibration. Unfortunately, it is common for modern altitude-azimuth telescopes with multiple mirror reflections to scramble the input polarization of the light before it reaches the analyzer of a coud\'{e}, Nasmyth, or Gregorian polarimeter. Solar astronomers in particular have developed a repretoire of techniques for calibrating telescopes (\citealt{elm92, kuh93, gir03, soc05a, soc05b, soc05e, sel05, soc06, ich08, kel09, elm10}. Our experience with the Advanced Electro-Optical System (AEOS) 3.67m telescope is that linearly polarized light can even, in some circumstances, be transformed to nearly 100\% circularly polarized light. There are 7 highly oblique reflections before polarization is analyzed at our High Resolution Visible and Infra-red Spectropolarimeter (HiVIS) (cf. \citealt{hod00}, \citealt{tho03}, \citealt{har06, har09, har10}, \citealt{har07, har08, har09a, har09b}). 	

	Effective calibration of the telescope and instrument minimizes these systematic errors and enhances the polarimetric precision of an observation. Moving optical elements, detector effects such as non-uniform sensitivity or nonlinearity, atmospheric seeing, and transparency variations can all induce polarization errors that are often larger than the photon noise. Strategies to minimize, stabilize and calibrate the instrument and telescope often involve use of non-moving optics and emphasize rapid chopping and common-path differential techniques. Calibration techniques include observing unpolarized and polarized standard stars, and use of polarized optical calibration units that are designed to mimic the nominal illumination of the telescope optics during target observations.

	The new technique presented here is designed to yield the full Mueller matrix calibration of many telescope-polarimeter systems with an accuracy of a few percent. It requires only two or more observations of the linearly polarized sky due to solar Rayleigh scattering. The daytime sky is normally easily observable, highly polarized, and is a relatively well characterized calibration source. It is observable without using precious dark observing time and requires no hardware calibration modifications of the telescope. As long as multiple linear polarization input states are detected at each telescope pointing, the full telescope Mueller matrix can be recovered with an accuracy of a few percent. This is realized by observing the solar illuminated sky with the Sun at different zenith and azimuth positions. This paper illustrates the technique using the Haleakala Air Force AEOS 3.7m telescope on Maui.

\subsection{Polarization} 

The following discussion of polarization formalism closely follows \citealt{col92} and \citealt{cla09}. In the Stokes formalism, the polarization state of light is denoted as a 4-vector:

\begin{equation}
{\bf S}_i = [ I,  Q, U, V ] ^T
\end{equation}

In this formalism, $I$ represents the total intensity, $Q$ and $U$ the linearly polarized intensity along polarization position angles $0^\circ$ and $45^\circ$ in the plane perpendicular to the light beam, and $V$ is the right-handed circularly polarized intensity. Note that according to this definition, linear polarization along angles $90^\circ$ and $135^\circ$ will be denoted as $-Q$ and $-U$, respectively. The typical convention for astronomical polarimetry is for the +$Q$ electric field vibration direction to be aligned North-South while +$U$ has the electric field vibration direction aligned to North-East and South-West.

The normalized Stokes parameters are denoted with lower case and are defined as:

\begin{equation}
[ 1,q, u, v ]^T = [ I,Q, U, V ] ^T / I
\end{equation}

The degree of polarization can be defined as a ratio of polarized light to the total intensity of the beam:
	
\begin{equation}
P = \frac {\sqrt{ Q^2 + U^2 + V^2 } } {I} =  \sqrt{ q^2 + u^2 + v^2 }
\end{equation}

	For details on polarization of light and stellar spectropolarimetry, see \citealt{col92} and \citealt{cla09}. 
	
	To describe how polarized light propagates through any optical system, the Mueller matrix is constructed which specifies how the incident polarization state is transferred to the output polarization state. The Mueller matrix is a $4 \times 4$ set of transfer coefficients which when multiplied by the input Stokes vector (${\bf S}_{i_{input}}$) gives the output Stokes vector (${\bf S}_{i_{output}}$):
	
\begin{equation}
{\bf S}_{i_{output}} = {\bf M}_{ij} {\bf S}_{i_{input}}
\end{equation}	

	If the Mueller matrix for a system is known, then one inverts the matrix and deprojects a set of measurements to recover the inputs.	 One can represent the individual Mueller matrix terms as describing how one incident state transfers to another. In this work we will use the notation:

\begin{equation}
{\bf M}_{ij} =
 \left ( \begin{array}{rrrr}
 II   	& QI		& UI		& VI		\\
 IQ 	& QQ	& UQ	& VQ		\\
 IU 	& QU	& UU	& VU		\\
 IV 	& QV		& UV		& VV		\\ 
 \end{array} \right ) 
\end{equation}

\subsection{Deriving Telescope Mueller Matrix Elements}

	Typical calibration schemes on smaller telescopes often use fixed polarizing filters placed over the telescope aperture to provide known input states that are detected and yield terms of the Mueller matrix. This approach has been used for solar telescopes (\citealt{soc05a, soc05b}, \citealt{soc05e, soc06}). Another approach, also used in the solar observations has been to image known sources and use spectropolarimetric data and Zeeman effect symmetry properties to isolate and determine terms in the Mueller matrix (\citealt{kuh93}, \citealt{elm10}). Night-time observations can use unpolarized and polarized standard stars to measure polarization properties of telescopes (cf. {\citealt{hsu82}, \citealt{sch92}, \citealt{gil03}, \citealt{fos07}). Many studies have either measured and calibrated telescopes, measured mirror properties or attempted to design instruments with minimal polarimetric defects (cf. \citealt{san91}, \citealt{gir03}, \citealt{pat06},\citealt{tin07}, \citealt{joo08}, \citealt{van09}, \citealt{roe10}). 

\subsection{The Polarized Sky as a Calibration}

	The observed polarization of the daytime sky is a useful calibration source. Scattered sunlight is bright, highly (linearly) polarized and typically illuminates the telescope optics more realistically than calibration screens. A single-scattering atomic Rayleigh calculation (cf. \citealt{cou80, cou98}) is often adequate to describe the skylight polarization. More realistic modeling, involving multiple scattering and aerosol scattering, is also available using industry standard atmospheric radiative transfer software packages such as MODTRAN (cf. \citealt{fet02},  \citealt{ber06}). 

\begin{figure}[h!,!t,!b]
\begin{center} 
\includegraphics[width=0.90\linewidth]{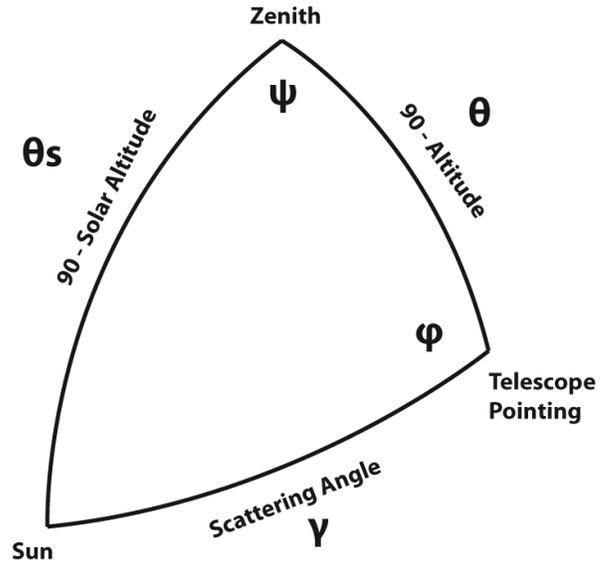} 
\caption{\small The celestial triangle representing the geometry for the sky polarization at any telescope pointing.  $\gamma$ is the angular distance between the telescope pointing and the sun. $\theta_{s}$ is the solar angular distance from the zenith. $\theta$ is the angular distance between the telescope pointing and the zenith. $\phi$ is the angle between the zenith direction and the solar direction at the telescope pointing. $\psi$ is the angle between the telescope pointing and the solar direction at the zenith. The law of Cosines is used to solve for any angles needed to compute degree of polarization and position angle of polarization.} 
\label{Rayleigh_Sky_Triangle} 
\end{center}
\end{figure} 

	There are many atmospheric and geometric considerations that determine the skylight polarization at a particular observatory site. The linear polarization amplitude angle can depend on the solar elevation, atmospheric aerosol content, aerosol vertical distribution, aerosol scattering phase function, wavelength of the observation and secondary sources of illumination (cf. \citealt{hor02a, hor02b}, \citealt{lee98}, \citealt{liu97}, \citealt{suh04}, \citealt{gal01}, \citealt{ver00}, \citealt{pom01}, \citealt{cro05, cro06}, \citealt{heg07}).  Anisotropic scattered sunlight from reflections off land or water can be highly polarized and temporally variable (\citealt{lit10}, \citealt{pel09}, \citealt{he10}, \citealt{sal07}, \citealt{ota10}, \citealt{kis04}). Aerosol particle optical properties and vertical distributions also vary (cf. \citealt{wu97}, \citealt{shu06}, \citealt{ver00}, \citealt{oug02, oug05a, oug05b, oug09a, oug09b}, \citealt{oug04}). The polarization can change across atmospheric absorption bands or can be influenced by other scattering mechanisms (cf. \citealt{boe06}, \citealt{zen08} \citealt{abe99, abe01}). 

	Deviations from a single scattering Rayleigh model grow as the aerosol, cloud, ground or sea-surface scattering sources affect the telescope line-of-sight. Clear, cloudless, low-aerosol conditions should yield high linear polarization amplitudes and small deviations in the polarization direction from a Rayleigh model. Observations generally support this conclusion (\citealt{pus05, pus06a, pus06b, pus07, pus08, pus09}, \citealt{sha10}).
	
	The geometry of our Rayleigh sky model is seen in Figure \ref{Rayleigh_Sky_Triangle}. The geometrical inputs are the observers location (latitude, longitude, elevation) and local time. The solar location and relevant angles from the telescope pointing are computed from the spherical geometry in Figure \ref{Rayleigh_Sky_Triangle}. The maximum degree of polarization ($\delta_{max}$) in this model occurs at a scattering angle ($\gamma$) of 90$^\circ$. The Rayleigh sky model predicts the degree of polarization ($\delta$) at any telescope pointing as:
	
\begin{equation}
\delta=\frac{\delta_{max} sin^2\gamma}{1+cos^2\gamma}
\end{equation}

	Since the angle of polarization in the Rayleigh sky model is always perpendicular to the scattering plane, one can derive the angle of polarization with respect to the altitude and azimuth axes of the telescope. The law of Cosines for the scattering plane breaks down at the Zenith where $\theta$ is 0$^\circ$ and one must simply calculate the difference in azimuth between the telescope pointing and the solar azimuth to find the orientation of the polarization. An example of the model we compared with our observations is shown in Figure \ref{lovis-rayleigh-sky} for Haleakala on January 27$^{th}$ 2010.

	One set of all-sky polarization measurements obtained at the Mauna Loa Observatory, a nearby site and at very similar altitude to Haleakala (\citealt{dah09}) are particularly relevant. The maximum degree of polarization ($\delta_{max}$) changed throughout the day in their observations from a value of $\sim$60\% with the sun at an elevation of 30$^\circ$, to 85\% with the sun setting. Although this is a significant change in the maximum degree of polarization ($\delta_{max}$) presumably due to multiple scattering, the polarization direction at all telescope pointings is expected to be close to the Rayleigh model (cf. \citealt{dah09}, \citealt{suh04}, \citealt{pus09}, \citealt{pom01}). In our analysis below we will use the Rayleigh model to define the input polarization angle. We believe our assumed input angle is accurate over most of the sky to $<$5$^\circ$  (eg. \citealt{suh04}).

\begin{figure} [!h, !t, !b]
\begin{center}
\includegraphics[width=0.85\linewidth, angle=90]{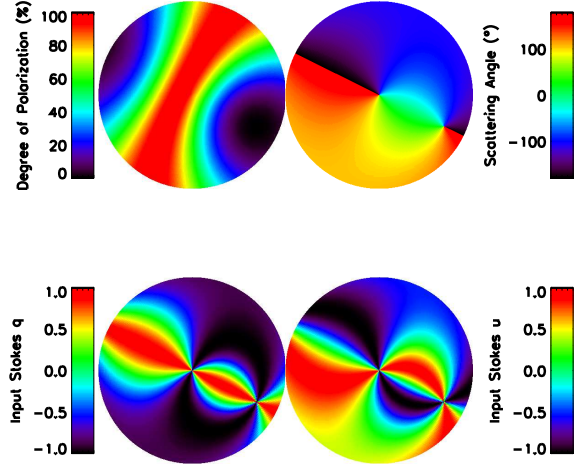}
\caption{ \label{lovis-rayleigh-sky} This Figure shows the sky polarization modeled for Haleakala at 3:00 UT on January 27$^{th}$ 2010. The degree of polarization ($\delta$) at all telescope pointings with $\delta_{max}$ scaled to 100\% is in the top left panel. The scattering angle with respect to the sun is in the top right panel. The solar altitude is 15$^\circ$ and the solar azimuth is 243$^\circ$. The calculated $q$ and $u$ inputs as seen by the AEOS altitude-azimuth telescope are on the bottom two panels. +$Q$ is defined as E-field vibrations along the +Altitude direction. +$U$ is defined as E-field vibrations along the +Altitude, +Azimuth direction. All panels show stereographic projections to the observed sky with the Zenith in the center of the circle. North, East, South, West are on the perimeter of the circle with North up and West right.}
\end{center}
\end{figure}

\subsection{Using the Sky for Full-Stokes Calibration}

	Despite the fact that the Rayleigh sky only provides linearly polarized light to the telescope, with reasonable assumptions, we can recover the full Mueller matrix.  We find that the telescope and spectrograph only weakly polarize unpolarized input light, and the optical system has only a weak depolarization effect on polarized input light. Thus, the first row and column of the system Mueller matrix can be approximated by the corresponding row and column of the identity matrix. Direct measurements of these terms using stellar observations with AEOS confirm that over the visible spectrum these Mueller matrix terms are less than 5\%. With this assumption the telescope system scrambles the input polarization simply by rotating the input 4 element Stokes vector in the 3-space defined along the $Q$, $U$, and $V$ polarization axes to yield the output measured state. 

	In general only two measurements with different input linear polarization orientations are sufficient to determine the telescope polarization properties. In practice we define a least-squares problem that takes several input polarization measurements of the sky at different times but with identical optical configurations  to derive telescope properties.

\section{Polarized Sky Observations}

	 To demonstrate this technique, we collected sky observations with AEOS using the new low spectral resolution mode for our coud\'{e} spectrograph we call LoVIS. This new spectrograph has been characterized to verify the expected performance using standard stars, calibration optics and various tests outlined in the Appendix. The LoVIS system polarization properties most relevant to this work include the telescope and instrument induced polarization, depolarization, the polarimetric response of LoVIS. These are briefly described below and in the appendix.
	
\begin{figure} [!h, !t, !b]
\begin{center}
\includegraphics[width=0.7\linewidth, angle=90]{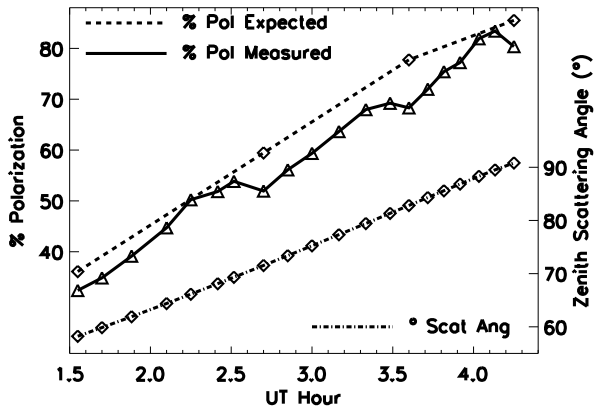}
\caption{ \label{lovis-127-degpol} This Figure shows the average sky polarization with the telescope pointed at the Zenith measured by LoVIS on January 27$^{th}$ 2010 UT. The statistical error in the measurement is much smaller than the width of the line. The solid line is the average polarization measured for all wavelengths. The dashed line is a Rayleigh Sky model prediction for the polarization at the Zenith following measurements from Mauna Loa in \citealt{dah09} Figure 7. The maximum degree of sky polarization in the 90$^\circ$ scattering plane ($\delta_{max}$) is assumed to rise linearly from 65\% to 85\% as the sun sets from elevation 30$^\circ$ to 0$^\circ$. The scattering angle ($\gamma$) between the sun and the Zenith telescope pointing is shown in the right hand y-axis.}
\end{center}
\end{figure}

	The measured degree of polarization for the many LoVIS sky observations we have obtained shows that there is relatively low instrument depolarization. For example, on January 27$^{th}$ 2010 UT we have twenty sky polarization observations with the telescope pointed at the Zenith taken over three hours ending just after sunset.  The measured degree of polarization averaged over all wavelengths is shown in Figure \ref{lovis-127-degpol} as the solid line. Following \citealt{dah09} Figure 7, we assume the maximum degree of polarization ($\delta_{max}$) for the Rayleigh sky as roughly 65\% when the solar elevation is 30$^\circ$, 75\% at 15$^\circ$ and 85\% at 0$^\circ$. If we apply this simple linear relation for $\delta_{max}$ and compute the Rayleigh sky polarization ($\delta$) at the Zenith for our observations, we get the dashed curve in Figure \ref{lovis-127-degpol}. The Figure also shows the scattering angle ($\gamma$) between the Zenith telescope pointing and the sun.
	
	The computed degree of polarization changes with wavelength for all twenty Jan 27$^{th}$ observations is shown in Figure \ref{lovis-twipol}. The average polarization shown in Figure \ref{lovis-127-degpol} was removed from the spectra leaving only residual chromatic  changes. There is a mild trend with wavelength for the measured polarization to decrease by about 10\% from 5500{\AA} to 7500{\AA}. The atmospheric oxygen absorption band shows a mild increase in polarization. There is also a substantial change seen the chromatic trend seen in the final observation taken just after sunset as shown by the dashed line of Figure \ref{lovis-twipol} consistent with studies of the occultation and changing illumination (cf. \citealt{oug02, oug05a, oug05b, oug09a, oug09b}, \citealt{oug04}).

	Given that AEOS telescope and LoVIS spectropolarimeter induces polarization of less than 3\% and the depolarization is of the same order, we can use a $3 \times 3$ Mueller matrix approximation. In this $3 \times 3$ $QUV$ space, the degree of polarization is preserved for any input state and the 3x3 Mueller matrix would be approximated as a rotation matrix inside the Poincar\'{e} sphere.

\begin{figure} [!h, !t, !b]
\begin{center}
\includegraphics[width=0.7\linewidth, angle=90]{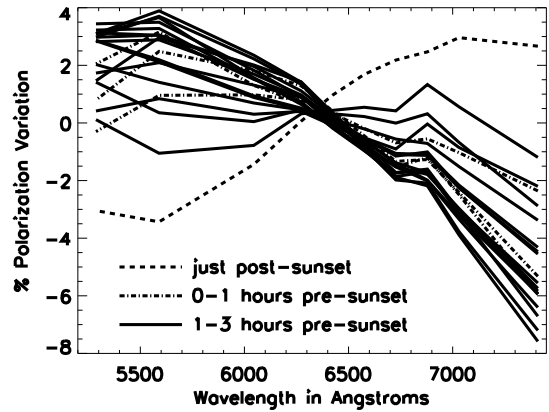}
\caption{ \label{lovis-twipol} This Figure shows the continuum polarization variation from average as a function of wavelength for all twenty LoVIS Zenith measurements on Jan. 27$^{th}$ 2010. The average value of the polarization, shown in Figure \ref{lovis-127-degpol} has been removed from all observations to highlight the change with wavelength. The solid lines show observations taken 2-3 hours before sunset. The dot-dash lines show observations taken 0.2-1 hour before sunset. The dashed line shows a single observation taken just after sunset.}
\end{center}
\end{figure}

\section{Telescope Polarization As Rotation}

	To model the the $3 \times 3$ Mueller matrix as a rotation matrix, we will use Euler angles.  We scaled all our measured Stokes vectors to unit length in order to remove the residual effects from changes in the sky degree of polarization, telescope induced polarization and depolarization; and to put our measurements on the Poincare spher\'{e}. We denote the 3 Euler angles as ($\alpha, \beta, \gamma$) and use a short-hand notation for sines and cosines where $cos(\gamma)$ is shortened to $c_\gamma$. We specify the rotation matrix (${\bf \mathbb{R}}_{ij}$)  using the ZXZ convention as:

\begin{equation}
{\bf \mathbb{R}}_{ij} =
 \left ( \begin{array}{rrr}
 c_\gamma	& s_\gamma	& 0		\\
 -s_\gamma	& c_\gamma	& 0		\\
 0			& 0			&1		\\ 
 \end{array} \right ) 
 \left ( \begin{array}{rrr}
 1			& 0			& 0		\\
 0			& c_\beta		& s_\beta	\\
 0			& -s_\beta		& c_\beta	\\ 
 \end{array} \right ) 
 \left ( \begin{array}{rrr}
 c_\alpha		& s_\alpha	& 0		\\
 -s_\alpha		& c_\alpha	& 0		\\
 0			& 0			&1		\\ 
 \end{array} \right ) 
\label{eqn_definerot}
\end{equation}

	With this definition for the rotation matrix, we solve for the Euler angles assuming a fully linearly polarized daytime sky as calibration input. If we denote the measured Stokes parameters, ${\bf S}_i$, as ($q_m, u_m, v_m$) and the input sky Stokes parameters, ${\bf R}_j$, as ($q_r, u_r, 0$) then the $3 \times 3$ $QUV$ Mueller matrix elements at each wavelength are:

\begin{equation}
{\bf S}_i =
\left ( \begin{array}{r}
q_{m} \\
u_{m} \\
v_{m} \\
\end{array}  \right ) =
{\bf M}_{ij} {\bf R}_j = 
 \left ( \begin{array}{rrr}
 QQ  &  UQ &  VQ     \\
 QU  & UU  & VU         \\
 QV  &  UV  &  VV     \\ 
 \end{array} \right ) 
\left ( \begin{array}{r}
q_{r} \\
u_{r} \\
0 \\
\end{array}  \right )
\label{eqn-rayltransf}
\end{equation}

	This set of equations has 6 variables and only 5 known quantities. We have no $V$ input to constrain the $VQ$, $VU$ and $VV$ terms. Nevertheless, two input and output vectors on the Poincare sphere are sufficient to fully specify the full $3 \times 3$ rotation matrix.  Thus, we use the fact that the sky polarization changes orientation with time and take measurements at identical telescope pointings separated by enough time for the solar sky illumination to change. This yields an over-constrained solvable problem for all six linear polarization terms in the Mueller matrix. 

\begin{figure} [!h, !t, !b]
\begin{center}
\includegraphics[width=0.7\linewidth, angle=90]{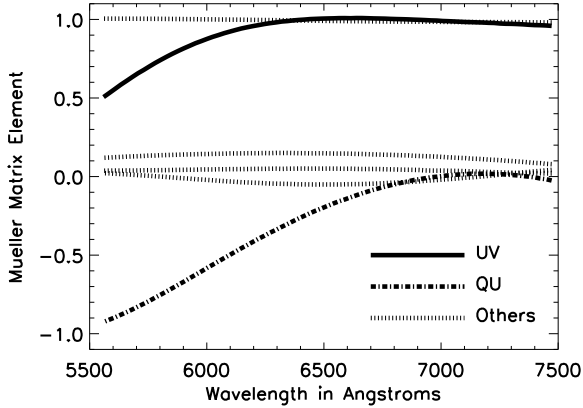}
\caption{ \label{lovis-mmat_elem} This Figure shows the six Mueller matrix elements versus wavelength at a telescope pointing of 90$^\circ$ altitude, 225$^\circ$ azimuth. These were derived from three sky measurements on January 27$^{th}$ 2010 using the two-step solution. The $UV$ term grows with increasing wavelength while the $QU$ term decreases with wavelength. The other Mueller matrix elements remain near 0 or 1 and show smaller chromatic variation.}
\end{center}
\end{figure}

	When using this rotation matrix approximation for the telescope Mueller matrix, the Rayleigh Sky input Stokes parameters multiply each term of the rotation matrix to give a system of equations for the three Euler anlges ($\alpha$, $\beta$, $\gamma$). This system of equations can be solved using a normal non-linear least-squares minimization by searching the ($\alpha$, $\beta$, $\gamma$) space for minima in squared error. This direct solution of this set of equations using standard minimization routines is subject to several ambiguities that affect convergence using standard minimization routines. For example, Euler angles are symmetric under the exchange of ($\alpha$, $\beta$, $\gamma$) with ( -180$^\circ$+$\alpha$, -$\beta$, -180$^\circ$+$\gamma$). Solutions are also identical with multiples of 2$\pi$. As a convenient approach to this least-squares problem, we used a different set of equations in a two-step solution. We first perform simple least squares solution directly for the six linear-polarization Mueller matrix elements. In the second step we fit a rotation matrix to the six Mueller matrix elements at each wavelength. As an example, Figure \ref{lovis-mmat_elem} shows the Mueller matrix elements derived with the Jan. 27$^{th}$ 2010 observations. The details of our methods for deriving Euler angles and an example of how one could plan sky calibration observations are outlined in the Appendix.

\begin{figure} [!h, !t, !b]
\begin{center}
\includegraphics[width=0.7\linewidth, angle=90]{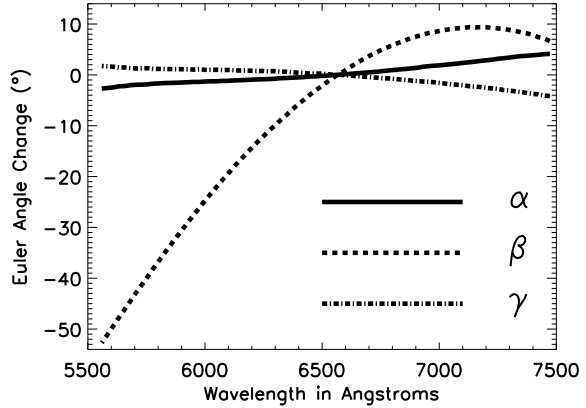}
\caption{ \label{lovis-rotmat-derived} This Figure shows the variation in Euler angles with wavelength at a telescope pointing of 90$^\circ$ altitude, 225$^\circ$ azimuth. The value for each Euler angle at 6560{\AA} have been removed for clarity [$\alpha$=176.5$^\circ$, $\beta$=81.6$^\circ$, $\gamma$=98.8$^\circ$]. The change with wavelength is dominated by $\beta$ which corresponds to the changes in the QU and UV Mueller matrix elements from Figure \ref{lovis-mmat_elem}. The statistical noise is smaller than the width of the line.}
\end{center}
\end{figure}

\subsection{Euler Angle Solution Properties}

	The Euler angles are well behaved smooth functions of both wavelength and telescope pointing. For instance, the Euler angle variations with wavelength calculated from January 27$^{th}$ 2010 observations at a telescope pointing of at 90$^\circ$ altitude and 225$^\circ$ azimuth are shown in Figure \ref{lovis-rotmat-derived}. The change with wavelength is dominated by $\beta$ while smaller variation is seen in $\alpha$ and $\gamma$. For clarity in the Figure, the Euler angle values at 6560{\AA} have been removed so all curves overlap. Similar curves are seen for other telescope pointings.

	The Euler angles are also smooth functions of telescope pointing. We have observations on multiple days taken on a grid of 8 azimuths and 3 different altitudes.  Figure \ref{eulers_vs_pointing} shows the derived Euler angles at 6560{\AA} as a functions of azimuth. There are three different curves corresponding to the three different telescope altitudes. We find that $\gamma$ effectively absorbs the changing azimuth while $\alpha$ strongly varies with altitude. Since there is substantial telescope induced rotation, these curves  show significant deviations from a purely geometrical rotation. 
	
	The derived Euler angles in Figure \ref{lovis-rotmat-derived} are quite repeatable when derived from calibration observations taken on different days. For instance, we have two sets of sky observations taken December 10$^{th}$ and 11$^{th}$ 2009 each consisting of thee observation sets spaced roughly one hour apart. The Euler angles derived from on different days agree to within 0.5$^\circ$ giving an estimate of the systematic error limits. The Mueller matrix elements derived with observations from different days agree to better than 0.01. The changes in $\beta$ with telescope azimuth shown in Figure \ref{eulers_vs_pointing} are well above the systematic noise of roughly 0.5$^\circ$ and are repeatable even when using many different combinations of calibration data on both days.  Similar trends for Euler angle variation with telescope pointing are seen at other wavelengths. 

\begin{figure} [!h, !t, !b]
\begin{center}
\includegraphics[width=0.7\linewidth, angle=90]{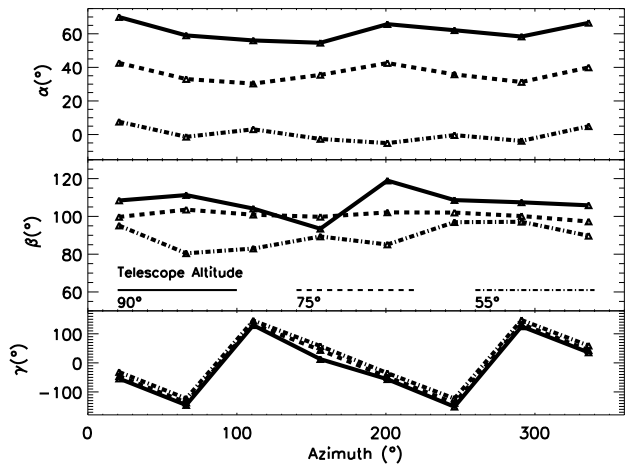}
\caption{ \label{eulers_vs_pointing} This Figure shows the variation in derived Euler angles at 6560{\AA} as a function of telescope pointing. The top panel shows $\alpha$ for all telescope azimuths at elevations of 90$^\circ$ (solid), 75$^\circ$ (dashed) and 55$^\circ$ (dot-dashed). The middle panel shows $\beta$ and the bottom panel shows $\gamma$ for all azimuths at the same elevations as $\alpha$ using the same line scheme. This Figure clearly shows that $\alpha$ changes the strongest with elevation while $\gamma$ shows changes with azimuth. There is not much change seen in $\beta$.}
\end{center}
\end{figure}

\subsection{Calibrating Spectropolarimetric Observations}

	Once telescope calibrations have been derived, observations subsequently taken with the instrument at the same telescope pointing can be de-rotated. The calibrated polarization measurements are then oriented in a reference frame on the sky corrected for all geometric, telescope and instrument induced rotation. We have done many experiments where sky observations on some days are used to calibrate sky observations taken on other days to examine the stability and repeatability of the calibrations.
	
	As an example, Figure \ref{lovis_meas_021} shows six individual sky spectropolarimetric measurements taken on December 10$^{th}$ and 11$^{th}$ 2009 at different times of the afternoon. All observations were taken at a telescope pointing of azimuth 021$^\circ$ and altitude 70$^\circ$. The input linear polarization orientation changes with time and there is significant chromatism and cross-talk observed in all Stokes parameters. The three observations from each day are taken at similar times and the observed polarization spectra are similar.

\begin{figure} [!h, !t, !b]
\begin{center}
\includegraphics[width=0.65\linewidth, angle=90]{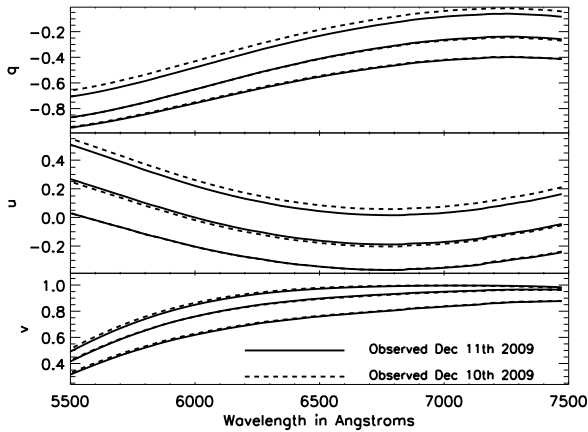}
\caption{ \label{lovis_meas_021} This Figure shows the measured spectropolarimetry with LoVIS while the telescope is fixed at altitude 70$^\circ$ and azimuth 021$^\circ$ scaled to 100\% polarization. Three observations were taken December 10$^{th}$ 2009 (solid lines) and another three observations were taken at similar times on December 11$^{th}$ 2009 (dashed lines). The statistical noise is smaller than the width of the line.}
\end{center}
\end{figure}
	
	In order to illustrate a general method of telescope calibration, imagine a scenario where we derive a set of calibrations using observations from one day and then use this telescope calibration to de-rotate observations taken on another day at an identical telescope pointing. We will treat the December 10$^{th}$ 2009 observations as a calibration set and derive Euler anlges for all telescope pointings and wavelengths. With these Euler angles in hand, we will then treat observations taken on December 11$^{th}$ 2009 as our science targets.  The de-rotated science observations will be compared against the Rayleigh sky prediction. This scenario is exactly what a typical observer would perform during routine operations. The difference between the de-rotated December 11$^{th}$ observations and the expected sky input Stokes parameters are shown in Figure \ref{lovis_derot_021}. The residual error is typically less than 0.01 for any Stokes parameter showing that the de-rotated observations match the expected sky model with very high accuracy.

\begin{figure} [!h, !t, !b]
\begin{center}
\includegraphics[width=0.7\linewidth, angle=90]{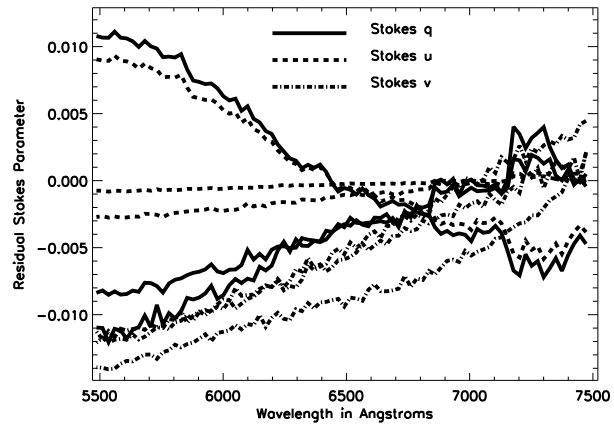}
\caption{ \label{lovis_derot_021} This Figure shows the difference between the de-rotated measured Stokes parameters and the predicted sky input Stokes parameters at altitude 70$^\circ$ and azimuth 021$^\circ$. The residual errors are less than 0.01 showing very low error between measured and expected Stokes parameters.}
\end{center}
\end{figure}

	We get consistent results for Euler angles using many different combinations of observations taken between August 2009 and March 2010.  Over this time period there were changes to the optical configuration such as remounting of the retarders, switching gratings and changing LoVIS / HiVIS modes as would be expected during normal observing operations. If one derives a residual rotation angle between the de-rotated measurements and the predicted sky, the residual rotation angles are almost always less than 3$^\circ$ when using any calibration set against any other observation set. If one uses calibration data where there are no optical changes between calibration and observation, this resulting angular error is much smaller, typically less than 0.5$^\circ$ rotation. This allows us to estimate systematic error from optical configuration changes as a few degrees residual rotation. When observing as calibrated, the errors are under 1$^\circ$ $QUV$ rotation.

\section{Discussion}

	We have demonstrated a simple method for deriving telescope polarization properties and Mueller matrix elements. By using observations of the bright, highly polarized daytime sky and taking observations at multiple times using identical optical configurations one can extract the polarization response of the telescope. 
	
	This method can be easily implemented via least squares methods. A direct least-squares solution for Euler angles and a two-step solution for the linear-polarization sensitive Mueller matrix elements and a rotation matrix fit to these elements have been implemented. Examples of how to optimally prepare calibration observations and solve for the telescope response properties are detailed in the Appendix.

	The AEOS telescope and LoVIS spectropolarimeter have over 20 reflections, many at high incidence angles before the polarization analyzer. However, the major polarization cross-talk effect is a simple rotations in $QUV$ space that is readily derived to an accuracy of a few percent from sky observations. Observations taken on multiple days show we can de-rotate LoVIS sky observations to the predicted Rayleigh sky with at least 0.5$^\circ$ accuracy.  

	The polarized sky is a calibration source that is bright, easily observable, highly polarized, simple to model, well characterized and easy to verify with additional instrumentation. Unpolarized and polarized standard stars have several disadvantages as calibration sources. Stars are not available at all telescope pointings and require using night-time observing time for calibration. Telescopes with complex optical trains such as AEOS require calibration at many pointings making the unavailability of stars at all pointings a serious problem. Solar instruments are typically unable to observe stars at all but can easily utilize this sky-based technique. Polarized standard stars normally have low degrees of polarization which adds time or noise constraints to such calibration efforts.  This technique uses sky-light to illuminate the telescope and instrument optics like the target observations. Night-time observations benefit from the fact that day-time observations are used to calibrate, rather than using precious night telescope time for calibration. 

\clearpage

\section{Appendix: Solutions for Euler Angles}

	There are a number of methods available to solve for Euler Angles and Mueller matrix elements given a set of sky polarization observations. Suppose we have measurements at different times but at identical telescope pointings. Multiplying out terms in equation \ref{eqn_definerot} for the ZXZ-convention we get: 
	
\begin{equation}
{\bf \mathbb{R}}_{ij} =
 \left ( \begin{array}{ccc}
 c_\alpha c_\gamma - s_\alpha c_\beta s_\gamma		& s_\alpha c_\gamma + c_\alpha c_\beta s_\gamma	& s_\beta s_\gamma		\\
-c_\alpha s_\gamma - s_\alpha c_\beta c_\gamma 		&-s_\alpha s_\gamma + c_\alpha c_\beta c_\gamma 	& s_\beta c_\gamma 	\\
 s_\alpha s_\beta								& -c_\alpha s_\beta								& c_\beta				\\ 
 \end{array} \right ) 
\label{eqn_rotmat_multout}
\end{equation}

Then equating Mueller matrix elements to rotation matrix elements could write the system of equations for the three Euler angles as:

\begin{equation}
\label{eqn_directsolve}
\begin{scriptsize}
\left ( \begin{array}{r}
q_{m1} \\
u_{m1} \\
v_{m1} \\
q_{m2} \\
u_{m2} \\
v_{m2} \\
q_{m3} \\
u_{m3} \\
v_{m3} \\
\end{array}  \right ) =
 \left ( \begin{array}{lll}
 q_{r1}(+c_\alpha c_\gamma - c_\beta s_\alpha s_\gamma) 	&+& u_{r1} (+c_\gamma s_\alpha + c_\alpha c_\beta s_\gamma)		\\
 q_{r1}(-c_\beta c_\gamma s_\alpha - c_\alpha s_\gamma) 	&+& u_{r1} (+c_\alpha c_\beta c_\gamma - s_\alpha s_\gamma)		\\
 q_{r1}(+s_\alpha s_\beta)							&+& u_{r1} (-c_\alpha s_\beta)								\\ 
 q_{r2}(+c_\alpha c_\gamma - c_\beta s_\alpha s_\gamma) 	&+& u_{r2} (+c_\gamma s_\alpha + c_\alpha c_\beta s_\gamma)		\\
 q_{r2}(-c_\beta c_\gamma s_\alpha - c_\alpha s_\gamma) 	&+& u_{r2} (+c_\alpha c_\beta c_\gamma - s_\alpha s_\gamma)		\\
 q_{r2}(+s_\alpha s_\beta)							&+& u_{r2} (-c_\alpha s_\beta)								\\ 
 q_{r3}(+c_\alpha c_\gamma - c_\beta s_\alpha s_\gamma) 	&+& u_{r3} (+c_\gamma s_\alpha + c_\alpha c_\beta s_\gamma)		\\
 q_{r3}(-c_\beta c_\gamma s_\alpha - c_\alpha s_\gamma) 	&+& u_{r3} (+c_\alpha c_\beta c_\gamma - s_\alpha s_\gamma)		\\
 q_{r3}(+s_\alpha s_\beta)							&+& u_{r3} (-c_\alpha s_\beta)								\\ 
 \end{array} \right ) 
\end{scriptsize}
\end{equation}

	This system of equations can be solved using a normal non-linear least-squares minimization by searching the ($\alpha$, $\beta$, $\gamma$) space for minima in squared error. With the measured Stokes vector (${\bf S}_i$), the Rayleigh sky input vector (${\bf R}_i$) and a rotation matrix (${\bf \mathbb{R}}_{ij}$) we define the error ($\epsilon$) as:

\begin{equation}
\label{eqn_rotmat_error}
\epsilon^2(\alpha,\beta,\gamma) = \sum \limits_{i=1}^{2} \sum \limits_{j=1}^{3} \left [ {\bf S}_i - {\bf R}_i {\bf \mathbb{R}}_{ij}(\alpha,\beta,\gamma) \right ]^2
\end{equation}

	For n measurements, this gives us $3 \times n$ terms in this least squares problem.  The Euler angles give identical rotation matrix elements under the exchange of ($\alpha$, $\beta$, $\gamma$) with ( -180$^\circ$+$\alpha$, -$\beta$, -180$^\circ$+$\gamma$) as well as with additional multiples of 2$\pi$. This direct solution is easily solvable in principle but the ambiguities make implementing this solution with existing software languages more cumbersome than necessary. 

	As an alternative method to the direct least-squares solution for Euler angles, we can do a two-step process that gives the same result but is much easier to implement. First we solve a system of equations for the Mueller matrix elements directly that are not subject to rotational ambiguity. With the Mueller matrix elements in hand, we can then perform a simple fit of a rotation matrix elements to the derived Mueller matrix terms. This two-step process allows us to have accurate starting guesses to speed up the minimization process, resolve Euler angle ambiguities and to estimate error propagation.

	When deriving the Mueller matrix elements of the telescope, one must take care that the actual derived matrices are physical. For instance, there are various matrix properties and quantities one can derive to test the physicality of the matirx (\citealt{giv93}, \citealt{kos93}, \citealt{tak09}). Noise and systematic uncertainty might give over-polarizing or unphysical Mueller matrices such as the element above that is $>$1. By simply using Mueller matrix elements to fit a rotation matrix, we avoid over-polarizing. 

	Provided the arrays are indexed properly, the normal solution for Mueller matrix elements can be computed via the normal least-squares method. We re-arrange the time-varying Rayleigh sky inputs to (${\bf R}_{ij}$) for $i$ independent observations and $j$ input Stokes parameters. The measured Stokes parameters (${\bf S}_i$) become individual column vectors. The unknown Mueller matrix elements are also arranged as a column vector by output Stokes parameter (${\bf M}_{j}$). If we write measured Stokes parameters as ($q_{m_i}, u_{m_i}, v_{m_i}$) and the Rayleigh input Stokes parameters as ($q_{r_i}, u_{r_i}$), we can explicitly write a set of equations for just two Mueller matrix elements:

\begin{equation}
{\bf S}_{i} =
\left ( \begin{array}{r}
q_{m_1} \\
q_{m_2} \\
q_{m_3} \\
\end{array}  \right ) = 
{\bf R}_{ij}{\bf M}_j =
\left ( \begin{array}{ll}
q_{r_1}	& u_{r_1}  \\
q_{r_2}	& u_{r_2}  \\
q_{r_3}	& u_{r_3}  \\
\end{array}  \right )
\left ( \begin{array}{r}
QQ	\\
UQ	\\ 
\end{array} \right ) 
\end{equation}

	We have three such equations for each set of Mueller matrix elements. With this can express the residual error (${\bf \epsilon}_i$) for each incident Stokes parameter (${\bf S}_i$) with an implied sum over $j$ as:

\begin{equation}
\label{eqn_mueller_lsq}
{\bf \epsilon}_i = {\bf S}_i - {\bf R}_{ij} {\bf M}_j
\end{equation}

	The normal solution of an over specified system of equations is easily derived in a least-squares sense using matrix notation. The total error $E$ as the sum of all residuals for $m$ independent observations we get:

\begin{equation}
E = \sum \limits_{i=1}^m {\bf \epsilon}_i^2
\end{equation}

	We solve the least-quares system for the unknown Mueller matrix element (${\bf M}_j$) by minimizing the error with respect to each equation. The partial derivative of Equation \ref{eqn_mueller_lsq} for ${\bf \epsilon}_i$ with respect to ${\bf M}_j$ is just the sky input elements ${\bf R}_{ij}$. Taking the partial with respect to each input Stokes parameter we get:

\begin{small}
\begin{equation}
\frac{\partial E}{\partial {\bf M}_j} = 2 \sum \limits_i {\bf \epsilon}_i \frac{\partial {\bf \epsilon}_i}{\partial {\bf M}_j} =  -2 \sum \limits_i {\bf R}_{ij} \left ( {\bf S}_i - \sum \limits_k {\bf R}_{ik} {\bf M}_k \right ) =  0
\end{equation}
\end{small}

We have inserted a dummy sum over the index $k$. Multiplying out the terms and rearranging gives us the normal equations:

\begin{equation}
\sum \limits_i \sum \limits_k {\bf R}_{ij} {\bf R}_{ik} {\bf M}_k  = \sum \limits_i {\bf R}_{ij} {\bf S}_i
\end{equation} 

Which written in matrix notation is the familiar solution of of a system of equations via the normal method: 

\begin{equation}
\label{eqn_muellerdemod}
{\bf M} = \frac{ {\bf R}^T {\bf S} } { {\bf R}^T {\bf R}}
\end{equation}

	This simple equation is very easy to implement with a few lines of code provided your observation times are chosen to give you a range of input states for a well-conditioned inversion. The noise properties and inversion characteristics of this equation can be calculated in advance of observations and optimized. We can write the matrix ${\bf A}$ with an implied sum over $i$ observations for each term as:

\begin{equation}
\label{eqn_rayl_modmatref}
{\bf A} =  {\bf R}^T {\bf R} = 
\left ( \begin{array}{rr}
q_{r_i}q_{r_i}	& q_{r_i}u_{r_i}		\\
q_{r_i}u_{r_i}	& u_{r_i}u_{r_i}		\\
\end{array}  \right )
\end{equation}

The solution to the equations for the three sets of Mueller matrix elements can be written as:

\begin{equation}
{\bf M}_j= \left ( \begin{array}{r}
QQ	\\
UQ	\\
\end{array}  \right ) =   {\bf A}^{-1} 
\left ( \begin{array}{l}
q_{r_i} q_{m_i}  \\
u_{r_i} q_{m_i}  \\
\end{array}  \right )
\end{equation}

\begin{equation}
{\bf M}_j= \left ( \begin{array}{r}
QU	\\
UU	\\
\end{array}  \right ) =   {\bf A}^{-1}
\left ( \begin{array}{l}
q_{r_i} u_{m_i}  \\
u_{r_i} u_{m_i}  \\
\end{array}  \right )
\end{equation}

\begin{equation}
\label{eqn-rayldemod}
{\bf M}_j= \left ( \begin{array}{r}
QV	\\
UV	\\
\end{array}  \right ) =   {\bf A}^{-1}
\left ( \begin{array}{l}
q_{r_i} v_{m_i}  \\
u_{r_i} v_{m_i}  \\
\end{array}  \right )
\end{equation}

As an example, if we compute the inverse of {\bf A} and multiply out ${\bf A}^{-1}$ for the $QQ$ term we can write:

\begin{equation}
\label{eqn-qq-solve}
QQ = \frac { (q_{r_i} q_{m_i})(u_{r_i} u_{r_i}) - (u_{r_i} q_{m_i})(q_{r_i}u_{r_i})}   { (q_{r_i}q_{r_i})(u_{r_i}u_{r_i})-(q_{r_i}u_{r_i})(q_{r_i}u_{r_i}) }
\end{equation}

In this manner, we can easily implement the usual matrix formalism with a time-series of daytime sky observations to measure six Mueller matrix elements. 

With the Mueller matrix terms in hand, solving for accurate initial Euler angle guesses is straightforward. As an example, by using $QV$ and $UV$ from \label{eqn_rotmat_multout} one can solve for $\alpha$ and $\beta$ directly:

\begin{equation}
tan(\alpha) = - \frac{QV}{UV}
\end{equation}

\begin{equation}
sin(\beta) = \frac{QV}{sin(\alpha)} = \frac{-UV}{cos(\alpha)}
\end{equation}

If one writes $c_\gamma$ as x then uses coefficients $a= \frac{QQ}{s_\alpha s_\beta}$ and $b= \frac{c_\alpha}{s_\alpha s_\beta}$  one can numerically solve a quadratic for and estimate of $\gamma$:

\begin{equation}
0 = (b^2+1)x^2 -(2ab)x + (a^2-1)
\end{equation}

The sign ambiguity in solving for $\gamma$=cos$^{-1}$(x) is resolved by comparing the sign of the computed Mueller matrix elements with those of the rotation matrix derived with these estimated Euler angles. The estimate shown here only utilized $QQ$, $QV$ and $UV$ but provided a guess for minimization routines typically accurate to better than 2$^\circ$ in our data set. 

In our two-step method, we next solved for the best fit Euler angles by doing a normal non-linear least-squares minimization by searching the ($\alpha$, $\beta$, $\gamma$) space around our initial guesses for minima in the summed squared error: 

\begin{equation}
\label{eqn_mmat_error}
E(\alpha,\beta,\gamma) = \sum \limits_{i=1}^{2} \sum \limits_{j=1}^{3} \left [ {\bf M}_{ij} - {\bf R}_{ij}(\alpha,\beta,\gamma) \right ]^2
\end{equation}

By using the built-in IDL routines POWELL or AMOEBA, we can construct the error function and find the minima. Both of these minimization routines require estimated starting points but give the same best-fit Euler angles to better than one part per million. Note that by choosing Euler angle guesses in the proper quadrants, the derived rotation matricies avoid the ambiguity of ($\alpha$, $\beta$, $\gamma$) with ( -180$^\circ$+$\alpha$, -$\beta$, -180$^\circ$+$\gamma$). The differences between rotation matrix elements and Mueller matrix elements for our observations typically much less than 0.1. The two-step solution gives us the same Euler angles as the direct least squares solution typically better than 0.2$^\circ$ and allows for computationally inexpensive and easy implementation of the least-squares minimization.

\begin{figure} [!h, !t, !b]
\begin{center}
\includegraphics[width=0.95\linewidth]{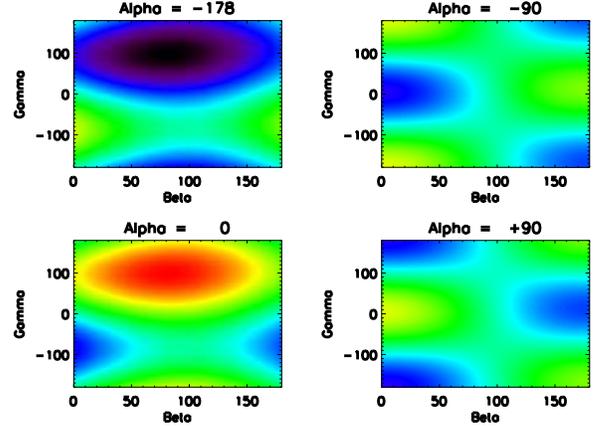}
\caption{ \label{error-function-dave} This Figure shows the squared error ($\epsilon^2$) for the two-step solution derived using Equation \ref{eqn_mmat_error}.  Each panel shows a slice through the volume ($\alpha$, $\beta$, $\gamma$) at four values of $\alpha$. Error increases from dark to light colors. Good solutions are found at $\alpha \sim 0$ and at $\alpha = \pm 180$.}
\end{center}
\end{figure}

In order to verify this method, we computed the error functions $E$ for both the direct solution and the two-step solution and verified the minima with a direct search of the entire Euler angle space. We created a grid of Euler angles from -180$^\circ$ to 180$^\circ$ (using 2 degree increments) and computed the error $E$ as in Equation \ref{eqn_rotmat_error}. Figure \ref{error-function-dave} shows the derived errors as functions of $\beta$ and $\gamma$ for four different values of $\alpha$. Each panel is color coded to show the square error $E$ increasing from dark to light. The Euler angles found using the two-step method are identical to those found with the direct solve within the crude 2$^\circ$ sampling we chose for this brute-force search. The error functions are always continuous with only two minima in the search volume showing convergence is easily achieved using any minimization scheme.

\clearpage

\section{Appendix: Planning Sky Observations Using Time Dependence for Modulation}

In order to use this technique efficiently, one must consider the noise propagation when inverting a sequence of sky observations to derive telescope Mueller matrix properties. There is an analogy between using the time-dependent Rayleigh sky to measure polarization properties of the a telescope and the retardances chosen to create an efficient modulation scheme for polarization measurements. The underlying mathematics is the same and one can easily derive the requirements on the observations needed to derive high-accuracy telescope Mueller matrix measurements. Effectively, one is attempting to derive properties of a matrix through what different groups call demodulations, inversions or deprojections. We outline here the analogy between polarimetric modulation and calculating noise properties with the Rayleigh sky to determine a quality observing sequence for deriving telescope Mueller matrices.

	Polarimeters modulate the incoming polarization state via retardance amplitude and orientation changes. This retardance modulation translated in to varying intensities using an analyzer such as a polarizer, polarizing beam splitter or crystal blocks such as Wollaston prisms or Savart plates. These modulation schemes can vary widely. For example, \citealt{com99},  \citealt{delToro00}, \citealt{dem03}, \citealt{nag07} and \citealt{tom10} overview optimal schemes, error propagation and outline schemes to maximize or balance polarimetric efficiency and create polychromatic systems. There have been many implementations of achromatic and polychromatic designs in both stellar and solar communities (cf. \citealt{gis03}, \citealt{han04}, \citealt{xu06}).  In the notation of these studies, the instrument modulates the incoming polarization information in to a series of measured intensities (${\bf I}_{i}$) for $i$ independent observations via the modulation matrix (${\bf O}_{ij}$) for $j$ input Stokes parameters (${\bf S}_j$): 

\begin{equation}
{\bf I}_{i} = {\bf O}_{ij} {\bf S}_{j}
\end{equation}	

This is exactly analogous to our situation where we have changed the matrix indices to be $i$ independent Stokes parameter measurements for $j$ different sky input Stokes parameters:

\begin{equation}
{\bf S}_{i} = {\bf R}_{ij} {\bf M}_{j}
\end{equation}	

In most night-time polarimeters, instruments choose a modulation matrix that separates and measures individual parameters of the Stokes vector: 

\begin{equation}
\label{normmod}
{\bf O}_{ij} =
 \left ( \begin{array}{rrrr}
 1   	& +1	&  0	&  0	\\
 1 	&  -1	&  0 	&  0	\\
 1 	&  0 	& +1	& 0	\\
 1 	&  0	& -1	&  0	\\ 
 1 	&  0	& 0	&  +1	\\ 
 1 	&  0	& 0	&  -1	\\ 
 \end{array} \right ) 
\end{equation}

Other instruments choose only four measurements: the minimum number of exposures required to measure the Stokes vector. In these schemes, one can balance the efficiency of the measurement to minimize the noise on each Stokes parameter: 

\begin{equation}
{\bf O} =
 \left ( \begin{array}{rrrr}
 1   	&  +\frac{1}{\sqrt{3}}	& +\frac{1}{\sqrt{3}}	&  -\frac{1}{\sqrt{3}}	\\
 1 	&  +\frac{1}{\sqrt{3}}	& -\frac{1}{\sqrt{3}} 	&  +\frac{1}{\sqrt{3}}	\\
 1 	&  -\frac{1}{\sqrt{3}} 	& +\frac{1}{\sqrt{3}}	&  +\frac{1}{\sqrt{3}}	\\
 1 	&  -\frac{1}{\sqrt{3}}	& -\frac{1}{\sqrt{3}}	&  -\frac{1}{\sqrt{3}}	\\ 
 \end{array} \right ) 
\end{equation}

One recovers the input Stokes vector from a series of intensity measurements by inverting the modulation matrix (${\bf O}$) provided it is a square and multiplying this inverse by the measured intensities. If the matrix is not square, one can simply solve the over-specified system of equations via the normal least squares formalism: 

\begin{equation}
\label{eqn_demod}
{\bf S} = \frac{ {\bf O}^T {\bf I} } { {\bf O}^T {\bf O}}
\end{equation}

Even for non-square matrices, we can define a demodulation matrix that captures the transfer properties of the modulation scheme:

\begin{equation}
{\bf D}_{ij} = [{\bf O}^T {\bf O}]^{-1} {\bf O}^T 
\end{equation}

In our case, the Rayleigh sky input parameters become the modulation matrix (${\bf O}_{ij} = {\bf R}_{ij}$) and the formalism for noise propagation developed in many studies such as \citealt{delToro00} apply.

As an example, we will use the Jan. 27$^{th}$ 2010 Mueller matrices derived in Figure \ref{lovis-mmat_elem}. The sky polarization rotated by about 33$^\circ$ during this period and the three calculated Rayleigh sky inputs scaled to 100\% polarization are:

\begin{equation}
\label{rayl_127_inputs}
\bf{R}_{ij} =
\left ( \begin{array}{ll}
q_{r_1}	& u_{r_1}  \\
q_{r_2}	& u_{r_2}  \\
q_{r_3}	& u_{r_3}  \\
\end{array}  \right ) =
\left ( \begin{array}{rrr}
+0.980	& +0.201	\\
+0.847	& +0.534	\\
 +0.716	& +0.698	\\
\end{array} \right ) 
\end{equation}

If each measurement has the same noise $\sigma$ and there are $n$ total measurements then the noise on each demodulated parameter (${\bf \sigma}_i$) becomes: 

\begin{equation}
\label{eqn_sigmas}
{\bf \sigma}_{i}^2 = n \sigma^2 \sum\limits_{j=1}^{n} {\bf D}_{ij}^2
\end{equation}

And the efficiency of the observation becomes:

\begin{equation}
\label{eqn_effics}
{\bf e}_{i} = \left ( n  \sum\limits_{j=1}^{n} {\bf D}_{ij}^2 \right ) ^{-\frac{1}{2}}
\end{equation}

For instance, the normal modulation sequence of equation \ref{normmod} used by most night time spectropolarimeters gives n=6 and ${\bf e}_i^2 = [1, \frac{1}{3}, \frac{1}{3}, \frac{1}{3}]$. The efficiency balanced scheme gives the same relative efficiencies as the normal modulation scheme but uses only 4 exposures instead of 6. In the case of our Jan. 27$^{th}$ observations considering only $qu$ terms our demodulation matrix is: 

\begin{equation}
{\bf D}_{ij} = [{\bf R}^T {\bf R}]^{-1} {\bf R}^T  =
\left ( \begin{array}{rrr}
+1.23	& +0.16	& -0.48	\\
-1.49		& +0.43	& +1.54	\\
\end{array} \right ) 
\end{equation}

With this demodulation matrix the efficiency for $u$ is only 60\% worse than $q$ and we have efficiencies of ${\bf e}_i = [0.43, 0.26]$ computed with Equation \ref{eqn_effics}. The Mueller matrix derived from these Rayleigh sky observations will have similar noise properties between $Q$ and $U$ terms.

One must take care with this technique to build up observations over a wide range of solar locations so that the inversion is well conditioned. The path of the sun throughout the day will create regions of little input sky Stokes vector rotation causing a poorly constrained inversion with high condition number. For instance, at our location in the tropics the sun rises and sets without changing azimuth until it rises quite high in the sky. We are constrained to observing in early morning and late evening with the dome walls raised since we may not expose the telescope to the sun. This causes input vectors at east-west pointings to be mostly $q$ oriented with little rotation over many hours. Observations at other times of the year or at higher solar elevations are required to have a well conditioned inversion.  One can easily build up the expected sky input polarizations at a given observing site with the Rayleigh sky polarization equations. Then it is straightforward to determine the modulation matrix and noise propagation for a planned observing sequence to ensure a well-measured telescope matrix with good signal-to-noise.

\clearpage

\section{Appendix: LoVIS Spectropolarimeter Characterization}

	In order to expand the capabilities of our spectropolarimeter, we have adapted the optics for use at low spectral resolution. The polarimeter unit for the spectrograph is a Savart plate providing dual-beam analyzing mounted behind the entrance slit with either two rotating achromatic retarders or two liquid-crystal variable retarders (LCVRs) outlined in \citealt{har08, har10}. We have modified the HiVIS echell\'{e} housing to allow a flat mirror to be mounted bypassing the echell\'{e} without disrupting the optical path. We call this new mode LoVIS.

\begin{figure} [!h, !t, !b]
\begin{center}
\includegraphics[width=0.7\linewidth, angle=90]{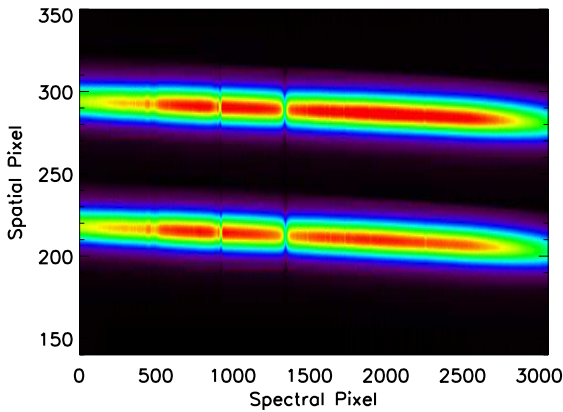} 
\caption{ \label{lovis_raw} This Figure shows example LoVIS data when observing a spectrophotometric standard star on Sept. 4$^{th}$ 2009. Wavelength increases from right to left. The Na D lines, H$_\alpha$ and the atmospheric A-band can be seen by eye near spectral pixels 2300, 1400 and 900 respectively.}
\end{center}
\end{figure}

	With a flat mirror replacing the echell\'{e}, the only dispersive element in the spectrograph is the cross-disperser working in 1$^{st}$ order. Only a single order is imaged on the detector making the illuminated region of the device much smaller. Figure \ref{lovis_raw} shows a raw data frame from a spectrophotometric standard star (HR7940) taken with LoVIS in spectropolarimetric mode using the Apogee detector. There is a single spectral order seen with orthogonally polarized spectra being displaced roughly 70 pixels spatially by the Savart plate. Wavelength increases from right to left with the H$_\alpha$ line seen near spectral pixel 1400 and the 6870{\AA} atmospheric band seen near pixel 900. We have two separate cross-dispersers available on the rotation stage blazed for 6050{\AA} and 8600{\AA} ( \citealt{hod00}, \citealt{tho03}). 
	
	The new LoVIS mode gives us much higher sensitivity and allows for much faster calibrations at spectral resolutions of 1,000 to 3,000 depending on the slit. The 1.5`` slit had spectral resolutions of 850 and 990 derived from Thorium-Argone (ThAr) lines at 5700{\AA} and 7000{\AA}. These lines were spectrally sampled with roughly 9.5 pixels per Gaussian Full-Width-Half-Max at each wavelength. The IDL reduction scripts we wrote for HiVIS outlined in \citealt{har08} were adapted to this new low-resolution data. Figure \ref{lovis_extract} shows the extracted spectrum from the raw data of Figure \ref{lovis_raw}. 

\begin{figure} [!h, !t, !b]
\begin{center}
\includegraphics[width=0.7\linewidth, angle=90]{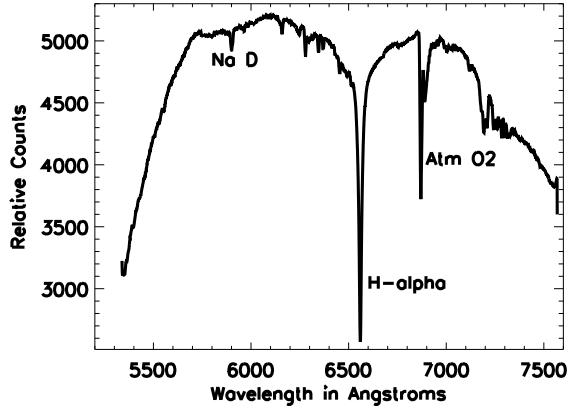}
\caption{ \label{lovis_extract} This Figure shows an example extracted spectrum of the spectrophotometric standard star HR7950 taken on Sept. 4$^{th}$ 2009. The H$_\alpha$, Na D lines and the atmospheric A-band are identified.}
\end{center}
\end{figure}

	This change in resolution allows us to observe much fainter targets or observe sources at a much higher cadence. When observing the daytime sky with AEOS and LoVIS, we can achieve a spectropolarimetric precision of 0.1\% with a 5 second exposure time using the 1.5`` slit. The polarimetric images sequences finish in one minute and the bulk of the overhead is from rotating the achromatic wave plates in the typical 6-exposure sequence.

	To calibrate this new polarimetric mode and the telescope polarization we used LoVIS to gather a wide range of sky and calibration observations from August 2009 to February 2010. All observations were taken with the Apogee detector described in \citealt{har10}. We have sky observations taken solely at the Zenith as well as observations on a grid of altitudes and azimuths. On several occasions we did observations for 3 or 4 hours continuously before sunset to accumulate a wide range of polarization inputs at multiple pointings. 
	
	In order to decouple the LoVIS and AEOS polarization measurements, we did a number of tests with our polarization calibration unit mounted at the slit as well as at the entrance port to our instrument at the calibration lamp unit. We did tests both without and with the image rotator inserted in the the path. This image rotator is the source of most of the cross-talk in the LoVIS instrument. Removing it makes the polarization properties of LoVIS much more benign as shown in Figures 15 and 16 of \citealt{har10}. There is a transmissive window that separates the coud\'{e} rooms from the central optical path to the telescope. This window is directly after the last fold mirror of the telescope, m7, and can be set to either BK7 or infrasil. This window can also have a polarimetric effect on the beam and both windows were also tested. Effectively we find that the polarization properties derived for HiVIS are similar to LoVIS and that we can efficiently reproduce polarization measurements with both rotating achromatic wave plates and LCVR modulators.

\begin{figure} [!h, !t, !b]
\begin{center}
\includegraphics[width=0.7\linewidth, angle=90]{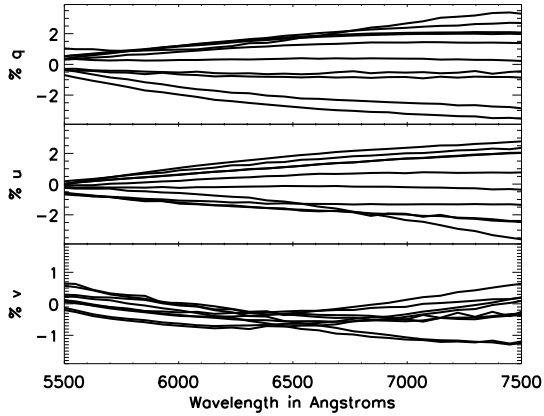}
\caption{ \label{lovis-stellar-continuum} This Figure shows the telescope induced polarization. We measured $quv$ spectra for many different unpolarized standard stars at many different pointings observed with LoVIS on Sept. 5$^{th}$ 2009. Each Stokes parameter is shown in percent. The induced polarization is at the few percent level with some dependence on wavelength.}
\end{center}
\end{figure}

	We also measured the induced polarization of the entire system when illuminated by a point-source utilizing unpolarized standard stars. As an example, we show observations of many unpolarized standard stars at many pointings from September 5$^{th}$ 2009 in Figure \ref{lovis-stellar-continuum}. The measured polarization is typical of results obtained on any individual night. Consistent with our previous HiVIS measurements from \citealt{har08}, the induced polarization is around a few percent, has some mild chromatic properties and does depend on telescope focus, the seeing and stellar location on the slit.

	In summary, the polarization properties of LoVIS are similar to HiVIS. The induced polarization is below a few percent. The polarization calibration optics give very pure inputs that are reproduced with both achromatic wave plate and LCVR modulators. This new low-resolution mode works efficiently and allows us to obtain calibration observations at a much faster cadence.

\end{document}